\documentclass[preprint, aps,amsmath,byrevtex,prd,titlepage, superscriptaddress,
tightenlines,
booktabs,
nofootinbib]{revtex4-1}

\usepackage{dcolumn}
\usepackage{amsmath}
\usepackage{amssymb}
\usepackage{bm}
\usepackage{hyperref}
\usepackage{graphicx}
\usepackage{epsfig}
\usepackage{slashed}

\usepackage{mathtools}

\newcolumntype{C}{>{$}c<{$}}

\newcommand{\beq}{\begin{equation}}
\newcommand{\eeq}{\end{equation}}
\newcommand{\eq}[1]{Eq.~(\ref{#1})}

\begin{document}

\title {New LHCb pentaquarks as Hadrocharmonium States}

\author{Michael I.~Eides}
\email[Email address: ]{meides@g.uky.edu}
\affiliation{Department of Physics and Astronomy, University of Kentucky, Lexington, KY 40506, USA}
\affiliation{Petersburg Nuclear Physics Institute, Gatchina, 188300, St.Petersburg, Russia}
\author{Victor Yu.~Petrov}
\email[Email address: ]{Victor.Petrov@thd.pnpi.spb.ru}
\affiliation{Petersburg Nuclear Physics Institute, Gatchina, 188300, St.Petersburg, Russia}
\author{Maxim V.~Polyakov}
\email[Email address: ]{maxim.polyakov@tp2.ruhr-uni-bochum.de}
\affiliation{Petersburg Nuclear Physics Institute, Gatchina, 188300, St.Petersburg, Russia}
\affiliation{Ruhr-University Bochum, Faculty of Physics and Astronomy,
Institute of Theoretical Physics II, D-44780 Bochum, Germany}


\begin{abstract}

New LHCb Collaboration results on pentaquarks with hidden charm \cite{Aaij:2019vzc} are discussed. These results fit nicely  in the hadrocharmonium pentaquark scenario \cite{Eides:2017xnt,Eides:2015dtr}. In the new data the old LHCb pentaquark $P_c(4450)$ splits into two states $P_c(4440)$ and $P_c(4457)$. We interpret these two almost degenerate hadrocharmonium states with $J^P=1/2^-$ and $J^P=3/2^-$ as a result of hyperfine splitting between hadrocharmonium states predicted in \cite{Eides:2015dtr}. It arises due to QCD multipole interaction between color-singlet  hadrocharmonium constituents.
We improve the theoretical estimate of hyperfine splitting \cite{Eides:2017xnt,Eides:2015dtr} that is compatible with the experimental data. The new $P_c(4312)$ state finds a natural explanation as a bound state of $\chi_{c0}$ and a nucleon, with $I=1/2$, $J^P=1/2^+$ and binding energy 42 MeV. As a bound state of a spin-zero meson and a nucleon, hadrocharmonium pentaquark $P_c(4312)$ does not experience hyperfine splitting. We find a series of hadrocharmonium states in the vicinity of the wide $P_c(4380)$ pentaquark that can explain its apparently large decay width. We compare the hadrocharmonium and molecular pentaquark scenarios and discuss their relative advantages and drawbacks.

\end{abstract}

\maketitle

\section{Introduction}

Pentaquarks with hidden charm were discovered by the LHCb Collaboration about five years ago \cite{Aaij:2015tga}. According to \cite{Aaij:2015tga}, there are two pentaquarks with hidden charm, one with a mass of $4380\pm8\pm29$ MeV and a width of $205\pm18\pm86$ MeV, and another with a mass of $4449.8\pm1.7\pm2.5$ MeV and a width of $39\pm5\pm19$ MeV. The preferred $J^P$ assignments are of opposite parity, with one state having spin 3/2 and the other 5/2. There is now an extensive theoretical literature on the interpretation of the LHCb pentaquarks, see, e.g., the recent  review \cite{Liu:2019zoy}. We will discuss the hadrocharmonium scenario, suggested in \cite{Dubynskiy:2008mq,Sibirtsev:2005ex,Li:2013ssa} (heavy quarkonium interaction with nuclei was considered in \cite{Brodsky:1989jd,Luke:1992tm}, see also references
in \cite{Voloshin:2007dx}). The hadrocharmonium approach to the LHCb pentaquarks was developed further in \cite{Kubarovsky:2016whd,Eides:2015dtr,Eides:2017xnt,Eides:2018lqg,Anwar:2018bpu,Perevalova:2016dln,Eides:2018lqg}. In our previous works \cite{Eides:2015dtr,Eides:2017xnt} we discussed interpretation of the LHCb pentaquarks as hadrocharmonium states, nonrelativistic bound states of $\psi(2S)$ and the nucleon. We described $P_c(4450)$ as a hadrocharmonium state \cite{Eides:2015dtr} with $I=1/2$, $J^P=3/2^-$, and calculated its decay widths \cite{Eides:2018lqg}. In the leading approximation the binding potential in hadrocharmonium does not depend on spin, so we predicted existence of a degenerate state with $I=1/2$, $J^P=1/2^-$. Degeneracy between the two color-singlet states with $J=1/2^-,3/2^-$ is lifted by a hyperfine interaction arising in the QCD multipole expansion, and the magnitude of the hyperfine splitting was estimated in \cite{Eides:2015dtr,Eides:2017xnt}.

New experimental data on the LHCb pentaquarks was presented recently  in \cite{Aaij:2019vzc}.  Two narrow states $P_c(4440)$ and $P_c(4457)$ are seen at the position of the old LHCb pentaquark $P_c(4450)$. Also a new narrow resonance $P_c(4312)$ shows up in the experimental data \cite{Aaij:2019vzc}.  We  consider discovery of two narrow states $P_c(4440)$ and $P_c(4457)$ as a confirmation of the prediction in \cite{Eides:2015dtr,Eides:2017xnt} of two almost degenerate pentaquark states with $J^P=1/2^-$ and $J^P=3/2^-$ and with the mass of the observed pentaquark $4450$~MeV.  We will improve the estimate \cite{Eides:2015dtr,Eides:2017xnt} of the hyperfine splitting between $J^P=1/2^-$ and $J^P=3/2^-$ pentaquarks below. We will use the approach developed in \cite{Eides:2018lqg} to calculate and compare partial and total decay widths of these pentaquarks. An interpretation of the new LHCb pentaquark $P_c(4312)$ as a hadrocharmonium bound state will be also discussed. The analysis in \cite{Aaij:2019vzc} was not sensitive to broad resonances, there was no new information on the status of the LHCb pentaquark $P_c(4380)$. We will elaborate on the hadrocharmonium scenario for this pentaquark \cite{Eides:2017xnt} below.

A natural theoretical framework for discussion of exotic mesons and baryons is provided by QCD at large $N_c$. It predicts a qualitative difference between exotic mesons and baryons. Light tetraquarks do not exist in QCD at large $N_c$ since meson-meson interaction decreases as $1/N_c$, see, e.g., \cite{sc1985}.  However, if quarks are so heavy that $m_Q\gg N_c\Lambda_{QCD}$ ($\Lambda_{QCD}$ is the scale of strong interactions), then  even a shallow potential $1/N_c$ can bind mesons with heavy quarks. The binding energy of such a molecular  state would be small since the binding potential is proportional to $1/N_c$. Therefore, only molecular type tetraquarks that are loosely bound states of two mesons with heavy quarks can exist in QCD at large $N_c$.

The case of exotic baryons is radically different. At large $N_c$ a baryon consists of $N_c$ quarks, its mass is proportional to $N_c$, and its interactions with mesons do not depend on $N_c$ at all. Hence, QCD at large $N_c$ bans existence neither of light nor heavy exotic baryons with the binding energy of order $\Lambda_{QCD}$. Unlike exotic mesons, large $N_c$ exotic baryons could be bound states of uniformly packed quarks having no resemblance to molecules, could have molecular structure \cite{Guo:2019fdo,Liu:2019tjn,Chen:2019bip,He:2019ify,Chen:2019asm,Fernandez-Ramirez:2019koa}, or have a more complicated hadrocharmonium structure.  The hadrocharmonium scenario makes unambiguous quantitative predictions, and  describes some fine features of the experimental data. Experimentally verifiable predictions of the hadrocharmonium scenario will be presented below.

\section{New LHCb Data and Hadrocharmonium Scenario}

The $P_c(4450)$ LHCb pentaquark was interpreted in \cite{Eides:2015dtr} as a hadrocharmonium bound state of $\psi(2S)$ and the nucleon. The binding potential in the hadrocharmonium picture is calculated using the QCD multipole expansion that holds only when the size of the charmonium state is smaller than the size of the nucleon. The multipole expansion is justified in the heavy quark and large $N_c$ limit, when the size of the nucleon is stable and the size of the charmonium excitation decreases with the mass of the heavy quark, see \cite{Eides:2015dtr,Eides:2017xnt} for more details. The leading contribution to the potential obtained in this way is proportional to the chromoelectric polarizability of the small color-singlet $c\bar c$ pair and is spin-independent. This spin-independence explains why the hadrocharmonium $P_c(4450)$ in \cite{Eides:2015dtr,Eides:2017xnt} is an almost degenerate doublet of states with spin-parities $J^P=1/2^-$ and $J^P=3/2^-$. As  shown in \cite{Eides:2015dtr,Eides:2017xnt}, the degeneracy between these color-singlet bound states is lifted by the hyperfine splitting  that arises due to interference of the chromoelectric dipole $E1$ and the chromomagnetic quadrupole $M2$ transitions in charmonium. Hyperfine interaction is described by the effective interaction Hamiltonian \cite{Voloshin:2007dx}

\beq \label{hyperfint}
H_{\rm eff} =-\frac{\alpha}{2m_Q}S_j\langle N(p')|E^a_i(D_iB_j)^a|N(p)\rangle,
\eeq

\noindent
where $E_i^a$ and $B_j^a$ are chromoelectric and chromomagnetic fields, and $S_j$, $\alpha$ and $m_Q$ are the $\psi(2S)$ spin, chromoelectric polarizability and the heavy ($c$) quark  mass, respectively.

The strength of the hyperfine interaction is determined by the chromoelectric polarizability and it is
additionally suppressed by the heavy quark mass $\sim 1/m_Q$ in comparison with the binding potential, see \cite{Eides:2015dtr,Eides:2017xnt} for more detail. Only the nucleon matrix element of the product of chromoelectric and chromomagnetic fields between the nucleon states with momenta $p$ and $p'$ in \eq{hyperfint} requires calculation. This matrix element can be estimated using the
approximations justified partially by theoretical and partially by experimental arguments (for more details see \cite{Balla:1997hf,Diakonov:1983hh,Diakonov:1985eg,Anthony:2002hy}). Then we obtain

\beq
\langle N(p')| E_i^a (D_i B_k)^a|N(p)\rangle \approx i q_i  \langle N(p')| E_i^a B_k^a |N(p)\rangle
\approx\frac{iq_k}{12}\langle N(p')|G^a_{\alpha\beta} \widetilde{G}^{a, \alpha\beta} |N(p)\rangle,
\eeq

\noindent
where $G^a_{\alpha\beta}$ are color field strengths and $q =p'-p$.

The flavor singlet axial current in QCD is anomalous that allows us to write the expression  on the right-hand side in terms of the singlet axial nucleon form factor $g^{(0)}_A(q^2)$

\beq
\langle N(p')| G^a_{\alpha\beta} \widetilde{G}^{a, \alpha\beta} |N(p)\rangle =\frac{32 \pi^2 }{ N_f} \ g_A^{(0)}(q^2)\  m_N  \bar u(p')i \gamma_5 u(p).
\eeq

\noindent
The hyperfine interaction Hamiltonian in \eq{hyperfint} reduces in the
coordinate space to the hyperfine potential for the $S$-wave hadrocharmonium bound state

\beq
\label{eq:Vhf}
V_{\rm hfs}({\bf r})=  \frac{g_A^{(0)}\alpha}{m_Q}  \frac{\pi M_A^4}{18 N_f }\  \frac{e^{-M_A r}}{r}\left(2 -M_A r\right)\ \left({\bf S\cdot s_N}\right),
\eeq

\noindent
where $\bm S$ and $\bm s_N$ are the spins of $\psi(2S)$ and the nucleon, respectively. We use the standard dipole  parameterization of the form factor $g^{(0)}_A(q^2)$ in terms of the form factor $g_A^{(0)}$ at zero momentum transfer and the dipole mass parameter $M_A$.

The value of $g_A^{(0)}\simeq 0.3$ can be obtained from the data on polarized deep inelastic scattering \cite{deFlorian:2009vb}. The dipole form factor is routinely used for description of all other nucleon form factors that are  measured experimentally. The value of the mass parameter $M_A$ is unknown but can be calculated in any nucleon model. The only real calculation which we are aware of is done in \cite{Silva:2005fa}. It produces $M_A=810$ MeV, see Table III in \cite{Silva:2005fa}. This result confirms the expectation that the mass parameter $M_A$ is determined by the radius of the nucleon and should be approximately the same in all channels. For example, it is well-known that in the vector channel it is also around $800$ MeV.

We used $\alpha=17.2$ GeV$^{-3}$ and hadrocharmonium wave functions from \cite{Eides:2015dtr,Eides:2017xnt} to calculate hyperfine splittings corresponding to different values of the dipole mass parameter $M_A$ in the interval $[0.8, 1.1]$~GeV. The results are presented in Table~\ref{tab:hfonMA}, so that the reader could compare predictions of his/her favorite nucleon model with the experimental data. We consider $M_A$ about 800 MeV as the preferred value of the mass parameter. Taking into account approximations employed in the calculations the expected accuracy of the mass splitting estimate is around 30\%. Comparison of the hyperfine splittings in Table~\ref{tab:hfonMA} with the experimental splitting between pentaquarks $P_c(4457)$ and $P_c(4440)$ shows satisfactory agreement between theory and experiment.

\begin{table}[h!]
\caption{Hyperfine mass splitting between $J^P=1/2^-$ and $3/2^-$ hadrocharmonium pentaquarks  as a function of the dipole mass parameter $M_A$   }
\begin{ruledtabular}
\begin{tabular}{ccccccc}
$M_A\ {\rm [GeV]}$ &  0.8 & 0.9&  1.0  & 1.1 \\
$\Delta E_{\rm hfs}\ {\rm [MeV]}$ & 21.1  & 27.7  & 34.9 & 42.5\\
\end{tabular}
\end{ruledtabular}
\label{tab:hfonMA}
\end{table}

Let us turn to the $P_c(4440)$ and $P_c(4457)$ decay widths. Partial decay widths of the hadrocharmonium and molecular pentaquarks  $P_c(4450)$ with $J^P=3/2^-$  were calculated in \cite{Eides:2018lqg}. We consider pentaquarks $P_c(4457)$ and $P_c(4440)$ as components of the hadrocharmonium hyperfine doublet and use old results for the hadrocharmonium with  $J^P=3/2^-$. In the same formalism as in \cite{Eides:2018lqg} we calculated now partial and total decay widths of the hadrocharmonium with $J^P=1/2^-$. All partial and total widths of both components of the hyperfine hadrocharmonium doublet are collected in Table~\ref{penthadrdec}. We see that decays to open charm of the $J^P=1/2^-$ hadrocharmonium state are enhanced. This happens because the partial wave with $l=0$ is allowed in these decays, to be compared with $l=2$ allowed in decays of the  $J^P=3/2^-$ hadrocharmonium. The central potential that contributes to the $l=0$ partial wave is stronger than the tensor potential that is responsible for the $l=2$ partial wave, for more details see \cite{Eides:2018lqg}. Additional accidental enhancement of $J^P=1/2^-$ decays is due to the larger Clebsch-Gordon coefficients in this decay.

The theoretical uncertainties of the total widths in Table~\ref{penthadrdec} are about 40\%, they are compatible with the experimental widths in \cite{Aaij:2019vzc} at the level of two standard deviations.  Experimentally the total width of $P_c(4440)$ is roughly more than three times larger than the width of $P_c(4457)$. Comparing with the theoretical results in Table~\ref{penthadrdec} we come to the conclusion that  $P_c(4440)$ has spin-parity $1/2^-$, while $P_c(4457)$ is a state with spin-parity $3/2^-$.

\begin{table}[h!]
\caption{\label{penthadrdec} Decay widths of $1/2^-$ and $3/2^-$  hadrocharmonium pentaquarks $P_c(4440)$ and $P_c(4457)$  }
\begin{ruledtabular}
\begin{tabular}
{lcc}
Decay mode
&
$\Gamma\left(\frac{1}{2}^-\right)$ [MeV]
&
$\Gamma\left(\frac{3}{2}^-\right)$ [MeV]
\\
\hline
$P_c\to J/\psi N$ &$11$      & $11$
\\
\hline
$P_c\to \Lambda_c\bar{D}$& $18.7$   &  $0.6$
\\
$P_c\to \Sigma_c\bar{D}$ & $1.4$  & 0.04
\\
$P_c\to \Lambda_c\bar{D}^*$ & $13.7$    & 4.2
\\
$P_c\to \Sigma^*_c\bar{D}$ & $0.004$    & 0.4
\\
\hline
Total width&44.8&16.2\
\end{tabular}
\end{ruledtabular}
\end{table}

The narrow LHCb pentaquark $P_c(4312)$ also finds a legitimate place in the hadrocharmonium scenario. We consider it as a bound state of the $\chi_{c0}(1P)$ charmonium state with $J^P=0^+$ and the nucleon. Interaction between $\chi_{c0}(1P)$ and the nucleon is determined by a symmetric  two-index chromoelectric polarizability tensor $\alpha_{ik}$ of $\chi_{c0}(1P)$. The effective $\chi_{c0}(1P)N$ interaction Hamiltonian has the form (see, e. g.,\cite{Voloshin:2007dx})

\beq \label{effectham}
H=-\frac{1}{2}\alpha_{ik}\langle N|E_i^aE_k^a|N\rangle.
\eeq

\noindent
The polarizability tensor can be written in the form

\beq
\alpha_{ik}=\alpha_1(J,S)\delta_{ik}+\alpha_2(J,S)J_iJ_k,
\eeq

\noindent
where $\bm S$ and $\bm J$ are the charmonium state spin and total angular momentum, respectively.

Then the $\chi_{c0}(1P)N$ interaction potential turns into a linear combination of a central and tensor potentials

\beq \label{nuclavforp}
H=V_c(r)+V_t(r)\left[(\bm n\cdot \bm J)(\bm n\cdot \bm J)-\frac{\bm J^2}{3}\right].
\eeq

\noindent
To calculate the potentials $V_c(r)$ and $V_t(r)$ one needs to find the average of the color-singlet operator $E^a_i E^a_j$ in the nucleon state, see \eq{effectham}. Numerous tensor structures could arise after this calculation. An estimate of the relative magnitudes of the arising potentials $V_c(r)$ and $V_t(r)$ can be obtained in the instanton approximation.   An effective approach to calculation of the hadron matrix elements of gluon operators in the instanton vacuum was developed in \cite{Diakonov:1995qy}. The most important feature of this approach is that integration over gluon fields reduces to  averaging over the instanton ensemble. Due to the hedgehog nature of instantons (rotations in color space are compensated by rotations in $SU(2)$ subgroup of $O(4)$) the color-singlet tensor $G^a_{\mu\nu}G^a_{\rho\lambda}$ reduces after integration over the color orientations of instantons to the combinations of Kronecker symbols and depends only on one invariant $G^2_{\mu\nu}$ which later can be expressed in terms of effective fermion operators. As a result, the second structure in the potential in \eq{nuclavforp} does not arise in the instanton vacuum and we omit the potential $V_t(r)$ in the estimates below.

\begin{table}[h!]
\caption{\label{pertpolp} Perturbative polarizabilities of heavy quarkonium $1P$ states in units of $\frac{a_B^4m_Q}{2N_c}$\footnote{$a_B$ is the quarkonium Bohr radius, $m_Q$ is the mass of the heavy quark, and $N_c$ is the number of colors.} }
\begin{ruledtabular}
\begin{tabular}
{ccccc}
$S$ & $J$ & $\alpha_1$ & $\alpha_2$ & $\alpha$
\\
\hline
$0$ & $1$ & $105$ & $-78$ & $53$
\\
$1$ & $2$ & $79$ & $-13$ & $53$
\\
$1$ & $1$ & $27$ & $39$ & $53$
\\
$1$ & $0$ & $53$ & $0$ & $53$
\end{tabular}
\end{ruledtabular}
\end{table}

The potential $V_c(r)$ in \eq{nuclavforp} differs from the $\psi(2S)N$ interaction  potential calculated in \cite{Eides:2015dtr,Eides:2017xnt} only by the value of the chromoelectric polarizability $\alpha=(1/3)\sum_i\alpha_{ii}$. Perturbative polarizabilities of the heavy Coulombic quarkonium $P$-states can be calculated in QCD perturbation theory similarly to the $S$-state calculations in \cite{Peskin:1979va,Bhanot:1979vb}. The results of perturbative  calculations are collected in Table~\ref{pertpolp}. Real charmonium is not a Coulombic bound state, so results of the perturbative calculations should be taken with a grain of salt. We expect that ratios of perturbative polarizabilities are closer to the real world than their absolute values. The ratio of perturbative polarizabilities for $2S$ and $1P$ states is $\alpha(1P)/\alpha(2S)=159/251\approx 0.63$. The Schr\"odinger equation for $\chi_{c0}(1P)$ and the nucleon has a bound-state solution with the experimental mass of the LHCb pentaquark $P_c(4312)$ when the interaction potential is 0.58 times weaker than in \cite{Eides:2015dtr,Eides:2017xnt}. Taking into account that polarizabilities are not Coulombic we consider the substitution $0.63\to0.58$ to be well inside the error bars of our calculations. Thus we identify the hadrocharmonium $\chi_{c0}(1P)N$ bound state with the LHCb pentaquark $P_c(4312)$, and predict that $P_c(4312)$ has spin-parity $1/2^+$. It does not have a hyperfine partner with approximately the same mass.

Let us discuss decays of the $P_c(4312)$ hadrocharmonium. Its total width about 10 MeV \cite{Aaij:2019vzc} can be easily explained as due to the decays of the weakly bound $\chi_{c0}(1P)$  that has total width of $10.8$ MeV dominated by decays into light hadrons. In addition, $P_c(4312)$ hadrocharmonium can decay into states with open charm. We expect that these decays are suppressed in comparison with such decays of the heavier pentaquarks (see Table~\ref{penthadrdec}) since the size of the hadrocharmonium $P_c(4312)$ is larger due to the smaller binding energy about 42 MeV to be compared with about 170 MeV for heavier hadrocharmonium states. All this does not explain the decay $P_c(4312)\to J/\psi+N$, where $P_c(4312)$ was observed. The parities of $\chi_{c0}(1P)$ and $J/\psi$ are opposite so transitional polarizability $\alpha(\chi_{c0}(1P)\to J/\psi)$ is zero and cannot explain this decay. The transition $\chi_{c0}(1P)\to J/\psi$ could go through exchange by three gluons, at least it is allowed by quantum numbers. An estimate of the hadrocharmonium pentaquark decay $P_c(4312)\to J/\psi+N$ is a challenging problem and we will not address it here.

\begin{table}[h!]
\caption{\label{psbound} Expected hadrocharmonium pentaquarks}
\begin{ruledtabular}
\begin{tabular}
{cccc}
Constituents  & Binding energy [MeV] & Mass [MeV] & Spin-parity
\\
\hline
$\eta_c(2S)N$ &  176.1  & 4401  & $1/2^-$  \\
$\chi_{c1}(1P)N$ &  44.2 & 4406  & $3/2^+,1/2^+$  \\
$h_c(1P)N$ &  43.9  & 4421  & $1/2^+,3/2^+$  \\
$\chi_{c2}(1P)N$ &  43.7  & 4452  & $5/2^+,3/2^+$  \\

\end{tabular}
\end{ruledtabular}
\end{table}

Hadrocharmonium interpretation of $P_c(4312)$ as a bound state of $\chi_{c0}(1P)$ and the nucleon naturally leads to the discussion of bound states of other charmonium $1P$ excitations and the nucleon. A trace of the polarizability tensor is one and the same for all $1P$ states, so the states $\chi_{c1}(1P)$, $\chi_{c2}(1P)$, and $h_c(1P)$ should also form bound states with the nucleon. In addition, the spin-zero  $S$-wave state $\eta_c(2S)$ should form a hadrocharmonium bound state with the nucleon  because its polarizability  coincides with that of  $\psi(2S)$. Solutions of the bound-state Schr\"odinger equations for all these states and their characteristics are collected in Table~\ref{psbound}. Minor differences between the binding energies of different $P$ states exceed the accuracy of our calculations and should be ignored.

We expect that degeneracy of the  states with the same spin will be lifted by hyperfine interaction, and the magnitude of this splitting will be roughly the same as the splitting between  $P_c(4440)$ and $P_c(4457)$. All charmonium constituents in Table~\ref{psbound} except $\eta_c(2S)$ have positive parity and natural widths about or below 1-2 MeV. We expect that decays of the type $(\chi_{c2}(1P)N)\to \chi_{c1}(1P)+N$ will go due to nonzero transitional polarizabilities $\alpha_{ik}(\chi_{c2}(1P)\to \chi_{c1}(1P))$ and have partial widths at the level of 10-20 MeV. Decays of the hadrocharmonium states in Table~\ref{psbound} to the states with open charm are also allowed and could have partial widths comparable with the ones for the decays to the states with hidden charm. Thus we expect that the interval of masses 4380-4430 MeV will be populated by a grid of hadrocharmonium states with the step 10-15 MeV and widths of order 10-30 MeV. We speculate that this set of states was interpreted in \cite{Aaij:2015tga} as a wide pentaquark $P_c(4380)$ and further experiment would resolve this structure in a series of relatively narrow overlapping resonances. Let us mention that the  $(\chi_{c2}(1P)N)$ hadrocharmonium state in the last line in Table~\ref{psbound} has a mass that almost coincides with those of the LHCb pentaquarks $P_c(4440)$ and $P_c(4457)$ that somehow makes this scenario less transparent.

\section{Summary}

We discussed above the hadrocharmonium interpretation of the new LHCb pentaquark results \cite{Aaij:2019vzc}\footnote{Recently the GlueX Collaboration reported  nonobservation of the $P_c(4450)$ pentaquark  in the photoproduction reaction $\gamma+p\to J/\psi+p$ \cite{Ali:2019lzf} that creates a certain tension between the LHCb and GlueX results, see, e.g., \cite{Cao:2019kst}. We expect that this nascent disagreement will be resolved in the near future.}.  The pentaquarks   $P_c(4440)$ and $P_c(4457)$ nicely fit the prediction \cite{Eides:2015dtr} of almost degenerate hadrocharmonium pentaquarks with $J^P=1/2^-,3/2^-$. We improved the estimate \cite{Eides:2015dtr,Eides:2017xnt} of hyperfine splitting between the color-singlet hadrocharmonium states due to interference of the $E1$ and $M2$ multipoles in the QCD multipole expansion (see, e.g., \cite{Voloshin:2007dx}) and obtained satisfactory quantitative agreement with the experimental data \cite{Aaij:2019vzc}, see Table~\ref{tab:hfonMA}.

We calculated partial and total widths of loosely bound hadrocharmonium  $(\psi(2S)N)$ states with  $J^P=1/2^-,3/2^-$ (see Table~\ref{penthadrdec}), and found that the total widths are compatible  with the experimental data for $P_c(4440)$ and $P_c(4457)$ \cite{Aaij:2019vzc}. Comparing the theoretical and experimental ratios of total widths we conclude that  $P_c(4440)$ has spin-parity $1/2^-$ and $P_c(4457)$ has spin-parity $3/2^-$.

The narrow LHCb pentaquark $P_c(4312)$ is naturally interpreted as a $(\chi_{c0}(1P)N)$ hadrocharmonium bound state with the binding energy 42 MeV, isospin $I=1/2$, and spin-parity $J^P=1/2^+$. Unlike the case of of almost degenerate hadrocharmonium $(\psi(2S)N)$ bound states with spin-parities $J^P=1/2^-,3/2^-$,  hadrocharmonium $\chi_{c0}(1P)N$  does not have a partner with another spin. This happens because $\chi_{c0}(1P)$ is a spin-zero state. We expect that the hadrocharmonium isodoublet pentaquark $P_c(4312)$ with $J^p=1/2^+$ has width about 10-20 MeV that arises due to natural decay width of the $\chi_{c0}(1P)$ charmonium and also due to open channels for decays into states with open charm.

We found a series of hadrocharmonium bound states with masses from 4380 MeV to 4430 MeV,  widths about 10-30 MeV and known spin-parities, see Table~\ref{psbound}. We speculate that these overlapping states were observed as a wide resonance $P_c(4380)$, and expect that future experiments will find a complicated structure in the vicinity of 4380 MeV.

Let us make a few remarks on the molecular scenario for the LHCb pentaquarks  $P_c(4440)$,   $P_c(4457)$, and $P_c(4312)$. This interpretation was suggested in \cite{Aaij:2019vzc}, and elaborated in a number of recent papers \cite{Guo:2019fdo,Liu:2019tjn,Chen:2019bip,He:2019ify,Chen:2019asm}. The molecular scenario is very attractive from the theoretical point of view and was originally suggested and developed for exotic hadrons in \cite{torn1991,teoegk1993,torn1994}. Exotic hadron in this approach is assumed to be a loosely bound state with a relatively large size. The binding potential between the two constituent hadrons is due to meson exchanges. The dominant role is played by the most long-range potential that is induced by the one-pion exchange. Like in the case of the deuteron, bound state arises as a result of interference between different partial waves, and its properties can be calculated almost without any additional assumptions, for more details see, e.g., \cite{torn1991,teoegk1993,torn1994,Eides:2017xnt}. A strong phenomenological argument in favor of this molecular picture of the new LHCb pentaquarks is their proximity to the two-particle thresholds of charmed particles. Two more massive pentaquarks  $P_c(4440)$ and $P_c(4457)$ are just below the $\Sigma_c^+\bar D^{*0}$ threshold 4460 MeV, and $P_c(4312)$ is just below the $\Sigma_c^+\bar D^{0}$ threshold 4318 MeV. Then the interpretation of $P_c(4312)$ as a bound state of $\Sigma_c^+\bar D^{0}$, and $P_c(4440)$ and $P_c(4457)$ as bound states of  $\Sigma_c^+\bar D^{*0}$ suggests itself.

On the other hand in the case of $P_c(4312)$ parity conservation bans  an effective  $\pi\bar{D}^0\bar{D}^0$ vertex and hence, the  pion exchange  does  not  give  any  contribution  in  the  binding  potential in the $\Sigma_c^+\bar D^{0}$ bound state. This observation makes interpretation of $P_c(4312)$ as a $\Sigma_c^+\bar D^{0}$ bound state less transparent because it is hard to understand how exchanges by heavier mesons that generate short-range potentials could be responsible for the existence of a loosely bound state with the constituents at relatively large distances, see also \cite{Fernandez-Ramirez:2019koa}. Emergence of a spin doublet of two narrow closely separated $P_c(4440)$ and  $P_c(4457)$ pentaquarks is also not quite natural in the molecular scenario.  Only one loosely bound $\Sigma_c\bar D^*$ state with $J^P=3/2^-$, $T=1/2$ arises if the principal contribution to the interaction potential is due to the long-range one-pion exchange, see, e.g., \cite{Eides:2017xnt}. Interaction in the $J^P=1/2^-$ channel is repulsive and there is no $\Sigma_c\bar D^*$ bound state with $J^P=1/2^-$. This difficulty can be resolved by considering a coupled-channel problem (with additional channels $\Sigma^*_c\bar{D}^*$, $\Sigma^*_c\bar{D}$, {\em etc}), see, e.g., \cite{He:2019ify,Chen:2019asm}. The threshold energies in the coupled channels are far from the bound states and the argument about closeness of the bound states to the two-particle thresholds is lost in this case. The coupled channels induce an effective attractive short range interaction and this additional short distance interaction generates binding in the $J^P=1/2^-$ channel that is necessary to fit the experimental data\footnote{See however, recent preprint \cite{Burns:2019iih}.}.

Both the hadrocharmonium and molecular scenarios have their advantages and drawbacks as we discussed above. The hadrocharmonium approach has effectively only one adjustable parameter (polarizability of the $2S$ charmonium state) and elegantly describes (really predicts, see \cite{Eides:2015dtr,Eides:2017xnt}) small mass splitting  between $P_c(4440)$ and $P_c(4457)$ as due to the QCD effective Hamiltonian in \eq{hyperfint}. It also uniquely predicts spin-parities of $P_c(4312)$, $P_c(4440)$, and $P_c(4457)$ and their decay widths. According to the hadrocharmonium scenario the mass interval 4380-4430 MeV is densely populated by hadrocharmonium resonances with widths of order 10-30 MeV. Predictions in the hadrocharmonium approach have a certain rigidity, they can be experimentally confirmed or falsified. For example, if it would turn out that the spin-parities of $P_c(4440)$ and $P_c(4457)$ are not $1/2^-$ and $3/2^-$, one will be compelled to abandon their interpretation as hadrocharmonium bound states $(\psi(2S)N)$. Molecular interpretation of pentaquarks is very flexible, due to the freedom to choose magnitudes of different coupling constants and parameters of numerous form factors it can accommodate almost any experimental data. This flexibility, that is advantageous in fitting the experimental data, weakens the predictive power of the molecular approach. For example, as we just discussed, the molecular scenario can describe  two closely separated narrow resonances $P_c(4440)$ and $P_c(4457)$ \cite{Chen:2019asm} observed by the LHCb Collaboration, but it fails to  give a natural explanation of their proximity to each other.

There is a significant number of experimentally verifiable predictions that are different in the hadrocharmonium and molecular scenarios. The quantum number assignments for the LHCb states do not coincide.  For example, parity of  $P_c(4312)$ is negative in the molecular picture \cite{Chen:2019asm} and it is positive in the hadrocharmonium one. One more way to test both models is to consider the decay patterns. We have calculated partial and full decay widths in both pictures (see Table~\ref{penthadrdec} and \cite{Eides:2018lqg}) and obtained  an intuitively appealing result that decays into states with hidden charm dominate for hadrocharmonium, while the molecule dominantly decays into states with open charm. Thus the decay patterns of molecular and hadrocharmonium pentaquarks are vastly different. At the present stage both the molecular and hadrocharmonium scenarios need further theoretical development, and we hope that the  dichotomy between them could be resolved by future experimental data.

\acknowledgments

We are grateful to Valery Kubarovsky for very useful discussions. M.~I.~Eides and V.~Yu.~Petrov have been supported by the NSF grant PHY-1724638. M.~V.~Polyakov has been supported by the BMBF grant 05P2018.

\end{document}